\documentclass[10pt]{article}

\usepackage{epsf,amsfonts,amssymb,epsfig,amsmath}
\usepackage{multirow}

\def\signed #1{{\leavevmode\unskip\nobreak\hfil\penalty50\hskip2em
  \hbox{}\nobreak\hfil(#1)%
  \parfillskip=0pt \finalhyphendemerits=0 \endgraf}}

\newsavebox\mybox
\newenvironment{aquote}[1]
  {\savebox\mybox{#1}\begin{quote}}
  {\signed{\usebox\mybox}\end{quote}}

\usepackage{amssymb,amsmath,mathbbol,mathrsfs}
\usepackage[usenames,dvipsnames]{color}
\usepackage{hyperref}
\usepackage{xcolor}
\usepackage{stmaryrd}
\usepackage{authblk}
\usepackage{framed}
\usepackage{empheq}
\usepackage{slashed}
\usepackage{hyphenat}
\usepackage{cite}

\usepackage{mdframed}

\usepackage{amsthm}
\theoremstyle{plain}

\newtheorem*{proposal*}{Proposal}

\makeatletter
\def\namedlabel#1#2{\begingroup
   \def\@currentlabel{#2}%
   \label{#1}\endgroup
}
\makeatother

\newcommand{\be}{\begin{equation}}
\newcommand{\ee}{\end{equation}}
\newcommand{\ben}{\begin{displaymath}}
\newcommand{\een}{\end{displaymath}}
\newcommand{\bea}{\begin{eqnarray}}
\newcommand{\eea}{\end{eqnarray}}
\newcommand{\bean}{\begin{eqnarray*}}
\newcommand{\eean}{\end{eqnarray*}}
\newcommand{\ba}{\begin{array}}
\newcommand{\ea}{\end{array}}
\newcommand{\bi}{\begin{itemize}}
\newcommand{\ei}{\end{itemize}}

\begin{document}

\begin{titlepage}

\vfill

\begin{center}
   \baselineskip=16pt
   {\Large\bf Causal Set Quantum Gravity and the Hard Problem of Consciousness}
  \vskip 1.5cm
Fay Dowker \\
     \vskip .6cm
            \begin{small}
      \textit{ 
       {Blackett Laboratory, Imperial College, Prince Consort Road, London, SW7 2AZ, UK}\\\vspace{5pt}
	{Perimeter Institute, 31 Caroline Street North, Waterloo ON, N2L 2Y5, Canada}\\\vspace{5pt}
              }
              \end{small}                      
               \end{center}
\vskip 2cm
\begin{center}
\textbf{Abstract}
\end{center}
\begin{quote}

I develop Rafael D. Sorkin's proposal that a partially ordered process of the birth of spacetime atoms in causal set quantum gravity  provides an objective physical correlate of our perception of time passing. I argue that one cannot have a fully objective, external picture of the birth process because the order in which the spacetime atoms are born is a partial order. I propose that live experience in causal set theory is an internal view of the objective birth process in which events that are neural correlates of consciousness occur.
In causal set theory, what ``breathes fire'' into a neural correlate of consciousness is that which breathes fire into the whole universe: the unceasing, partially ordered process of the birth of spacetime atoms. 

\end{quote}

\vfill
\end{titlepage}

\tableofcontents

\newpage

\section{Introduction}\label{sec:intro}

\begin{aquote}{Mephistopheles in ``Faust'' by Johann Wolfgang von Goethe, 1808} 
Grau, teurer Freund, ist alle Theorie,
Und grün des Lebens goldner Baum.\\
$[$Grey, dear Friend, all Theory is, And green Life's golden Tree.$]$
\end{aquote}

\begin{aquote}{Rafael D. Sorkin  \cite{Sorkin:2007qh} } Think of the causal set as an idealized growing tree [...] Such a tree grows at the tips of its many branches, and these sites of growth are independent of
one another.  [...The causal set] is ``growing at the tips'' but not in a synchronized manner with respect to any external time. There is no single ``now'' that
spreads itself over the entire process. \end{aquote}

\vskip2cm

In his  Autobiographical Notes \cite{Schilpp:1949}, Albert Einstein states his
``epistemological  credo'' which begins:

\begin{quote}
I see on the one side the totality of sense-experiences, 
on the other side the totality of the concepts and propositions that are laid down in books. The relations between the concepts and propositions among themselves are of a logical type, and the business of logical thinking is strictly limited to the making of the connection 
between concepts and propositions among each other according to firmly-set rules, which are the concern of logic. The concepts and propositions acquire ``meaning'' and ``content'', respectively,  only through their relation to sense-experiences. The connection (Die Verbindung) of the latter with the former is purely intuitive, not itself of a logical nature. The degree of certainty 
with which this relation or intuitive connection (intuitive Verkn{\"{u}}pfung) can be made, and nothing else, is what differentiates empty fantasy from scientific ``truth''.
\end{quote}

\noindent  But, you may say, isn't Einstein's epistemology of science, a {theory} of science and 
scientific progress  that  is \textit{written down in a book} and that includes a \textit{concept} of sense-experience. And doesn't the ``truth'' of this theory, by its own lights, have to be judged solely on whether it can be brought into coordination with sense-experience? And so on. To avoid circling down an endless rabbit hole of self-reference, one must acknowledge that Einstein's epistemology of science is not a theory in the Einsteinian sense  of having definite axioms and logical rules of inference that result in
propositions about the concepts within the theory. It is, rather, a heuristic. It may illuminate and help with the project of advancing our understanding of the world or it may obscure and hinder. And this judgment may be made by anyone who takes part in the struggle, personally, and perhaps differently in different contexts. 

Is Einstein's epistemology of science helpful in thinking about the Hard Problem of consciousness \cite{Chalmers:1995,Chalmers:1996}? Let us take what Einstein refers to as ``sense-experiences'' and, in a famous diagram (letter to Maurice Solovine dated 7th May 1952 in \cite{solovine}), as  ``immediate (sense) experiences'' (die unmittelbaren (Sinnen) Erlebnisse) to be amongst those conscious experiences referred to by the Hard Problem. Then the Hard Problem looks hard indeed by the light of Einstein's epistemology. For, to solve it one must create  theoretical concepts of sense-experiences and lay these down in books.
And then, in order that the theory may be distinguished from empty fantasy, one must make an intuitive connection between sense-experiences and propositions about the sense-experience-concepts in the theory with some degree of certainty, a  task that is bound to be haunted by spectres of circularity and lurking paradox. 
 
In this article I aim to show that there is a concept in theoretical physics that, though it can be stated---``the unceasing process of the partially ordered births of spacetime atoms''---cannot be fully captured, not in books, nor in a movie, nor even in thought, as it is \textit{objectively}.  To the extent that I achieve this aim, I will have demonstrated that Einstein's epistemology of science, heuristic though it is, is not flexible enough to accommodate all of physics and in the wiggle room that opens up, I suggest, we can make progress on the Hard Problem. Indeed, the partially ordered birth process within the causal set approach to the problem of quantum gravity has been proposed by Rafael D. 
Sorkin as a physical correlate of our subjective experience of time passing. Sorkin states \cite{Sorkin:2007qh}: 
 \begin{quote}
It [the example of sequential growth models for causal sets] 
even provides an objective correlate of our subjective perception of “time passing” in the
unceasing cascade of birth-events that build up the causal set, by “accretion” as it were. [Footnote:] The notion of ``accretive time'' that arises here seems close to that of C.D. Broad, and
also to that of the ``Vibhajavadin'' school within the Buddhist philosophical tradition.\footnote{ A interesting project for a scholar of Buddhism interested in physics would be to try to determine the ``Vibhajavadin'' view of time, perception and epistemology in detail by examining the earliest existing records.}
\end{quote}

Sorkin's paper is a treatise not on consciousness but on time. However, there is a striking similarity between aspects of the debate between two broad camps loosely labelled Being and Becoming on the question of the nature of time and aspects of the debate between two broad camps loosely labelled Sufficiency of Neural Correlate of Consciousness (NCC) and Insufficiency of Neural Correlate of Consciousness on the question of consciousness. In each case, the latter camp (Becoming or I-NCC) claims that something essential is missing from the former camp's expositions. 
In each case, members of the former camp (Being or S-NCC) claim that  the ``something essential'' is an illusion and/or a misapprehension. In this article I take the position that the two debates are the same debate and that \textit{our perception of time passing is the aspect of consciousness to which the Hard Problem pertains}.\footnote{I thank Huw Price for referring me to Chapter V of Arthur Eddington's Gifford Lectures in which Eddington makes a connection between consciousness and the perception of the passage of time: ``[...] if there is any experience in which this mystery of mental recognition can be interpreted as insight rather than image-building, it should be the experience of ``becoming.'' [....]  The view here advocated is tantamount to an admission that consciousness, looking out through a private door, can learn by direct insight an underlying character of the world which physical measurements do not betray''. (Chapter V ``Becoming'' in \cite{Eddington1928})}
 It follows from this position that Sorkin's proposal is a solution to the Hard Problem and in this article I will develop it 
in the following form:\\
\medskip

\begin{mdframed}
\textbf{Proposal}  (referred to henceforth in this article as ``the Proposal'') \\
\textit{The process of the partially ordered birth of the atomic-events that compose an NCC-event in causal set theory correlates with the subject {having} the corresponding conscious experience in real time.
}
\end{mdframed}

\medskip

The plan of the article is as follows. Section \ref{sec:events} introduces the concept of \textit{event} in the mathematics and physics of stochastic processes. A simple such process---the random walk on the integers---is used to illustrate the two theories, one dynamic and one static, with different physical content, that can be derived from the same mathematics.  The concept of event can be carried over from stochastic processes to formally deterministic theories such as General Relativity and 
 section \ref{sec:basics} discusses events in GR and the primitive point-events of which every event is composed and which are {partially} ordered by precedence.\\
 Section \ref{sec:discreteness} introduces the causal set approach to the problem of quantum gravity and fundamental discreteness of spacetime. In causal set theory (CST) spacetime is composed of discrete, indivisible \textit{spacetime atoms} and there are  dynamical models in which spacetime grows by the accretion of spacetime atoms that are born in a partially ordered process. I develop a three-way distinction between (i) an event-as-such, (ii) the occurrence of an event and (iii) an occurred-event in the context of the partially ordered process of the birth of spacetime atoms. 
I argue that, because the order of the births is partial and not linear, one can have an {external}, dynamic view of the process  only {if} that view is partly subjective. The Proposal is that one {can}, nevertheless, have an {objective} dynamic view of the process if one is oneself part of the physical system: conscious experience \textit{is} that {internal} view, that \textit{in-sight}. \\
 
Section \ref{sec:consciousness} puts the partially ordered birth process to work on the Hard Problem.  In subsection \ref{sec:qualities} I build confidence in the Proposal  by making connections between properties of the partially ordered birth process and reported qualities of conscious experience. 
In subsection \ref{sec:inlightof} I build further confidence in the Proposal  by 
throwing its light on some aspects of the Hard Problem.
In subsection \ref{sec:sv}  I compare and contrast the Proposal  to that of Marina Cortes, Lee Smolin and Clelia Verde \cite{pittphilsci19530}.

Throughout the article I will assume that there is no relevant quantum interference between the physical histories referred to.  Section \ref{sec:discussion} is a conclusion section and includes a brief discussion of quantum dynamics in causal set theory. 

\subsection{Objective and subjective}

 For the purpose of the article the terms  ``objective'' and ``subjective'' have meaning with respect to a particular physical theory, the theory under discussion whatever that is in context:\\

\noindent  \textit{Objective = physical in the theory.} \\
\noindent  \textit{Subjective = not physical in the theory but exists in the minds of scientists working with the theory.}  \\

\noindent In physics,  gauge independent concepts are objective and pure gauge concepts are subjective. Consider these examples from GR. The proper time measure along a timelike worldline, the volume measure of a 4 dimensional region of spacetime and the event horizon of a Schwarzschild black hole are objective. The Schwarzschild coordinate system for the Schwarzschild black hole spacetime is subjective. Sometimes a concept can have both objective and subjective content. For example the Schwarzschild metric in Schwarzschild coordinates contains both objective, viz. physical information and subjective, viz. unphysical information.  

\section{The concepts of process and event in physical theory} \label{sec:events}
Informally,  ``event'' means ``something that can happen'' in the physical theory one is working within. 
In the applied mathematics of stochastic processes, ``event''  has a precise  definition: an event is a subset of the set of all possible histories for the system at hand.\footnote{Strictly, an event is a \textit{measureable} subset of the set of all possible histories but I will elide this technical point in what follows.} This latter definition will be the meaning of the term ``event'' throughout this article and given the importance of this meaning I will illustrate it using the concrete example of a discrete stochastic process that is a simple random walk on the integers. 

 There are two ways to conceive of the random walk: 
 \begin{itemize}
 \item[A] as a  \textit{dynamic} process and 
 \item[B] as a  \textit{static} measure theory.
 \end{itemize}
 
 Let us say, just for definiteness, that the walker starts at the origin at $t=0$ and steps one place to the left or to the right at each discrete stage of the random walk. Let us call the position of the walker after $t$ steps $\gamma_t$. 
 In both A and B, one considers the set $\Omega$ of all possible completed histories of the walker i.e. all possible infinite sequences of integers $\{\gamma_t\} = \{\gamma_0, \gamma_1, \gamma_2 \dots \} $ where $\gamma_0= 0$ and $|\gamma_k - \gamma_{k+1}| = 1$.  An \textbf{event} is a  subset of $\Omega$. For example the event ``the walker is at $5$ at stage $13$" is the subset $E(5;13) := \{ \{\gamma_t\} \in \Omega\,|\, \gamma_{13} = 5\} $. 

In A there are transition probabilities for the walker to step to the left or to the right at each stage and these transition probabilities define the random walk. 
In B  there is a probability measure on the collection of all events.  The probability measure on the events can be calculated from the set of all the transition probabilities and vice versa: the same information is contained in the measure and in the transition probabilities.   

In physical theory A there is a process and the walker steps to the left or right at each stage of the process.
In physical theory B  one infinite completed ``Block Universe'' history $\{\gamma_t\}$  is chosen at random from $\Omega$ according to the probability measure on the collection of all events. That history is the world and each event either has occurred or has not occurred in the world. For example, if history $\{\Gamma_t\}$ is chosen we look at $\Gamma_{13}$. If it equals $5$ we say that the event 
$E(5;13)$ happened and if it does not equal $5$ then we say that the event $E(5;13)$ did not happen.  

 A and B  are different physically.  In A there is the dynamic \textit{process of stepping} and a physical order in time in which the steps occur. In B  there is no process, there is no stepping. 
  
 In A, because there is a process of stepping,  there is a {difference} between an event---a subset of $\Omega$---and its occurrence in the process. For example, $E(5;13)$ is a set of histories and in physical theory A the occurrence of $E(5;13)$ is the stepping of the walker to the integer $5$ (from $6$ or $4$) at stage $13$ of the process. In theory B  the process of stepping and the concept of the occurrence of an event  are absent; theory B is static and is a stochastic ``process'' in name only.  The Proposal is built on the dynamic conception of stochastic process and the difference between an event and the occurrence of the event will be crucial. This will be set out in detail later in the article in the context of causal set theory in section \ref{sec:dynevents}. For now, I just emphasise the contingent nature of events: an event may or may not happen. This is important because we have considered here an example of a process, the random walk,  in which the dynamics is stochastic but in formally  deterministic theories such as GR and classical mechanics, events are still contingent---they may or may not happen---due to uncertainty about initial and other external conditions.\\
   
\noindent \textit{Note}:  The physicist can have an {external} view of the world in both of these physical theories of the random walk. In theory B the external view is static; it is an infinite sequence of integers. In theory A the external view is dynamic; it is a movie of a walker stepping to the left and the right in a linearly ordered sequence of steps.

\section{Events in General Relativity}\label{sec:basics}

The Proposal hangs on two particular aspects of events in General Relativity, aspects that are inherited by events in Causal Set Theory: \\

 \noindent { i.}  Every event includes a spacetime substrate as the event's most fundamental physical component.\\
  \noindent  { ii.}  The order of precedence of the primitive subevents of an event is a {partial} order. \\

The following subsections expand on these aspects of events in GR.

\subsection{Events in GR include a spacetime substrate} 

According to GR the physical world is a 4 dimensional (4D) continuum spacetime, together with matter, if it is present, described by decorations on spacetime. Matter here refers to all  physical stuff apart from spacetime, such as the trajectories of atoms and the electromagnetic field.  
 Spacetime in GR is not an empty vessel, nor is it a mere background stage for what happens. In GR, spacetime is a 4D physical entity with a geometric structure and dynamical laws in which spacetime and matter interact with each other. Spacetime is the entity that GR is the physics of.

Let us consider for definiteness, and inspired by Robert Geroch \cite{Geroch:1978},  an event $E_{fire}$  that is the explosion of a firecracker.  From an old-fashioned perspective, a conceptual left-over from Newtonian physics, $E_{fire}$ has spatial and temporal location and spatial and temporal extension. In GR the concepts of  temporal and spatial location and extension of $E_{fire}$ lose their separate meanings and it is only physically meaningful to say that $E_{fire}$ has location-extension in 4D spacetime. $E_{fire}$'s location-extension is a 4D region of spacetime of finite 4-volume of the order of $10^{-3} m^3 s$, say,  and I will refer to this region of spacetime as the \textit{spacetime substrate} 
 $SS(E_{fire})$ of the event $E_{fire}$.

The spacetime substrate of an event in GR is its most fundamental physical component.  For, in the absence of matter there are still events. If $R$ is a 4D region of spacetime that is a vacuum solution of the Einstein equations then $R$ is, by itself, an event. In our terminology we say there is an event, $E$, such that  $SS(E)= R$ and there are no matter decorations. On the other hand, in the absence of a spacetime substrate, there is no event; there is nothing.

  \subsection{Subevents and point-events}\label{sec:subevents}

The explosion event $E_{fire}$ has subevents corresponding to subregions of its spacetime substrate. For example a subregion 
might correspond roughly to a subinterval of the full time duration of the explosion event. 
One of the axioms of GR is that spacetime is a continuum and one can continue considering smaller and smaller subevents by considering smaller and smaller subregions of spacetime---roughly, regions of shorter time duration and smaller spatial extension \cite{Geroch:1978}---until in the infinite limit, $SS(E)$ can be partitioned into all of its uncountably infinitely many points. I will refer to ``the  point-event $p$ in $SS(E)$'' as shorthand for ``the point-event whose spacetime substrate is the point $p$ in spacetime region $SS(E)$.''  The smallest sub-events of $E$ in GR are its {point-events}.

\subsection{Order}\label{sec:order} 

An event $E$ is more than merely the collection of its point-sub-events: to reconstruct $E$ from its point-sub-events, one needs to know the physical precedence order relation on the point-sub-events together with the spacetime volume measure of $SS(E)$ \cite{Malament:1977}. I will denote by the symbol $\prec$ the physical precedence order relation on point-events in GR.\footnote{The usual term for the precedence order relation $\prec$ in GR textbooks and papers is the \textit{causal relation}.}  Some axiomatic formulations of GR allow spacetimes in which there is no precedence order relation because the spacetimes contain closed causal curves.  In this article I will assume that the precedence order exists because it is a prediction of causal set theory that such spacetimes with closed causal curves are unphysical and do not belong to ``GR proper''. 

Given any two point-events, $p$ and $q$, in the spacetime substrate of an event $SS(E)$, either  $p \prec q$ ($p$ precedes $q$) or $q \prec p$ ($q$ precedes $p$)  or $p$ and $q$ have no order.  Phrases synonymous with  ``$p\prec q$''  are $p$  precedes $q$, $q$ is preceded by $p$, $p$ is before $q$ and $q$ is after $p$. Two point-events $p$ and $q$ that are unordered are not simultaneous; there is no physical
concept of simultaneity in GR. If two point-events $p$ and $q$ are unordered then no physical influence propagates from $p$ to $q$ nor from $q$ to $p$,  according to the principle of relativistic causality that holds in GR.

Consider now our firecracker explosion event. 
In the region $SS(E_{fire})$,  GR tells us what the order of precedence is.  
Assuming that gravitational interactions can be neglected  $SS(E_{fire})$ will be a region of 4D Minkowski space. 
In an {inertial coordinate} system
$\{x^0, x^1, x^2, x^3\}$,  if $p$ has coordinates $(p^0, p^1, p^2, p^3)$  and $q$ has coordinates $(q^0, q^1, q^2, q^3)$ then $p\prec q$ if and only if 
 $p^0 < q^0$ and $  (p^1 - q^1 )^2 + (p^2- q^2)^2 +  (p^3- q^3)^2 \le (p^0 - q^0)^2$. The units of the coordinates are chosen so that the speed of light equals 1 in these units.
 
\begin{figure}
  \centering
    {\includegraphics[width=\textwidth]{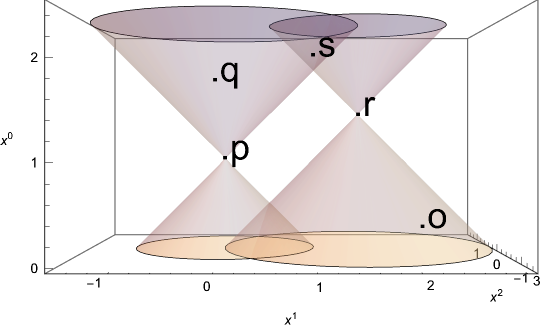}}
     \caption{The figure represents a subregion of $SS(E_{fire})$, a portion of 4D Minkowski spacetime with one coordinate, $x^3$ suppressed.  The $x^0$ coordinate along the vertical axis  is a timelike coordinate  and the other two axes are  spacelike coordinates.  The unit of coordinates $x^1$ and $x^2$ is $1cm$ and the unit of coordinate $x^0$ is the time taken for light to travel $1 cm$. The dots labelled $o$,
 $p$, $q$, $r$ and $s$  represent point-events in $SS(E_{fire})$. The solid cone above $p$
is composed of all point-events that are preceded by  $p$   and the solid cone below $p$ is composed of all point-events that precede $p$. The units are such that the aspect ratio of the cones equals one (the height equals the radius of the base). This order structure exists at every point-event  e.g.  the  precedence cones at $r$ are also shown. }
     \label{fig:mink}
\end{figure}

The structure of the order relation is important to understanding the Proposal  and a diagram will help guide our thoughts. Figure \ref{fig:mink} is a 3D cartoon  of the physical order of precedence of some example point-events of $SS(E_{fire})$. 
 The order of precedence for point-events  $o$, $p$, $q$, $r$ and $s$ in figure \ref{fig:mink} is: $o\prec r$, $r\prec s$, $o\prec s$, $p\prec q$, $p\prec s$ and there are no other order relations. $(p,r)$ is an example of a pair of point-events that have no order: neither precedes the other. 
This is all basic physics from GR and Special Relativity but it is worth reviewing given the central role that the particular, relativistic structure of the precedence order is going to play. 

It is hard to overstate how important the precedence order relation is in GR.  GR textbooks abound with conformal or Penrose-Carter diagrams that illustrate the precedence order. The epitome of GR, a black hole, is defined in terms of the precedence order. Many of the central theorems of GR are based on the techniques of global causal---viz. precedence---analysis pioneered by Roger Penrose.  
Also, when gravitational interactions can be neglected and the spacetime substrate is Minkowski spacetime---as it is for our example $E_{fire}$ event---or some other fixed spacetime, then we can consider quantum matter as a relativistic QFT in that spacetime. The relativistic structure of the precedence order then plays a crucial role in the construction of physical QFTs including the Standard Model of particle physics. 

\subsection{An external objective picture in GR}

A Block Universe is a maximal event, one whose spacetime substrate is maximally extended. A Block Universe is an external objective picture   of the entire history of the physical world in GR. It is an external picture, a view from the outside that can be shared and communicated. It is an objective picture, given in terms of physical concepts in the theory, spacetime, field configurations, particle trajectories etc. 

There is a huge literature on the question of whether the physical world in GR must be a Block Universe (see \cite{Putnam:1967, Rietdijk:1966} 
for arguments for the Block Universe in Minkowski spacetime,  as examples from this rich literature). GR with a Block Universe event as the physical world fits into Einstein's epistemology of science. It can be laid down in books. There is no concept of the physical passage of time in a Block Universe, however, and work must be done to establish an intuitive coordination between our perception of time passing and concepts within the Block  (see \cite{Williams1951-WILTMO-4, Dainton:2011, Ismael:2016} as examples from this literature).  Alongside such work, many suggestions have been made for re-envisioning GR and/or augmenting GR with additional structures so as to provide the theory with a physical passage of time (see for example \cite{10.2307/42705837} for an assessment of the possibility of reconciling Becoming with GR and \cite{Price:2011} for an assessment of alternatives to a Block Universe in general, not specific to GR). 

\section{Causal Set Theory}\label{sec:discreteness} 
 
Causal set theory (CST) is an approach to the problem of quantum gravity that  hypothesizes that 
spacetime is fundamentally discrete at the Planck scale and that maintains the centrality of the concept of precedence order. 
Spacetime in CST, then, takes the mathematical form of a discrete partial order or causal set, or causet for short  \cite{Myrheim:1978, tHooft:1979, Bombelli:1987aa, Surya_2019}.
Physically, a causet is a discrete set of indivisible spacetime atoms together with their physical order of precedence $\prec$, a partial order.

\subsection{Events in CST}\label{sec:cstevents} 

In CST it is supposed that an event such as our explosion event $E_{fire}$ in GR can be recovered from an event in CST, both kinematically and dynamically. These twin hypotheses---kinematical and dynamical recovery---are both needed for the Proposal. In this subsection the kinematics is discussed and in the next subsection \ref{sec:dynevents}, the dynamics is discussed. 

An event $E$ in GR is both its spacetime substrate $SS(E)$  and its matter content. In CST, the spacetime substrate $SS(E_{fire})$ is supposed to be recovered from, to be a large scale approximation to a finite causet $CS(E_{fire})$, as a fluid state in fluid mechanics is recovered from, is a large scale approximation to a discrete molecular state. There is good evidence for this hypothesis in CST (see the review in section 3 of \cite{Butterfield:2021}).   For example,  in the continuum the precedence order and the volume measure together are the full spacetime geometry \cite{Malament:1977} and 
it is plausible that the precedence order of the continuum is recovered from the precedence order of the underlying causet and that the volume measure of the continuum is recovered from the counting measure of the underlying causet.\footnote{If one is attracted to order as a fundamental concept then the promise of CST that spacetime is pure order is a powerful appeal.} 

 On the other hand, at the current stage of our knowledge in CST it is an almost completely open question how to account for matter. 
For the purposes of this article I will assume that the matter data of events such as $E_{fire}$ in GR can be recovered from some kind of decorations on the causet substrate $CS(E_{fire})$. 

Putting this together, and using the same notation $E_{fire}$ to denote the firecracker explosion event in CST, the kinematical hypothesis is:\\
\noindent The explosion event $E_{fire}$ in CST has a causet substrate $CS(E_{fire})$,  to which the continuum Minkowski spacetime substrate $SS(E_{fire})$ is a good approximation, together with decorations on $CS(E_{fire})$ for matter.\\

\noindent \textit{Note}: The event $E_{fire}$ in GR is recovered by many 
CST events that differ from each other on small scales, differences in both the causet substrate and the matter decorations that are irrelevant to $E_{fire}$ in GR. 
A reference to the event ``$E_{fire}$ in CST'' is a reference to any one of this large collection of CST events---it matters not which---each of which recovers the event $E_{fire}$ in GR. The fluid analogy is helpful. 
A particular fluid state at large scales is recovered by many discrete molecular states that differ from each other on small scales, differences that are irrelevant to the fluid. A reference to ``the molecular state'' of the fluid is a reference to any one of this large collection of molecular states---it matters not which---each of which recovers the fluid description.\\
 
In CST, unlike in the continuum, there is a finite end to the procedure of considering smaller and smaller subevents of $E_{fire}$ in CST: after a finite number of subdivisions one reaches indivisible \textit{atomic-events},  each with a single spacetime atom as its substrate. The explosion event $E_{fire}$ in CST  is {composed} of roughly $10^{144}$ indivisible atomic-events with their 
precedence order relations. That number is very large but it is finite and is set by the hypothesis that the scale of the discreteness in CST is the Planck scale.  Also, $E_{fire}$ in CST is \textit{nothing but} its atomic-events together with their
precedence order relations.  In GR, by contrast,  the precedence order relation is not
enough information to fully reconstruct an event $E$ from all its point-events: one also needs to know the volume measure on $SS(E)$ \cite{Malament:1977}.

Figure \ref{fig:mink} can now be reinterpreted as representing $E_{fire}$ in CST. The dots labelled $o$, $p$, $q$, $r$ and $s$ in figure \ref{fig:mink} can now be thought of as representing atomic-events embedded  in the approximating continuum Minkowski spacetime at the positions shown.  The physical order of precedence of the atomic-events $o$, $p$, $q$, $r$ and $s$ is the order of precedence their embedded positions have in the approximating Minkowski spacetime. So, $o\prec r$, $r\prec s$, $o\prec s$, $p\prec q$ and $p\prec s$,  and all other pairs of them have no order. $o$, $p$, $q$, $r$ and $s$ are but a tiny representative sample of the $10^{144}$ or so atomic-events composing 
$E_{fire}$, all of which should  be thought of as embedded uniformly throughout the approximating $SS(E_{fire})$.  There will be an even vaster, though still finite, number of precedence order relations.

\subsection{CST dynamics and three concepts pertaining to events}\label{sec:dynevents}

Causal set theory has produced novel dynamical models, the Classical Sequential Growth (CSG) models, in which a discrete universe grows 
in an unceasing process of the birth of spacetime atoms \cite{Rideout:1999ub}.\footnote{Christian W{\"u}thrich and Craig Callender make a critical assessment of CSG models using the concept of Becoming as a relation \cite{Wuthrich:2015vva}.} The birth process is the physical passage of time in CSG models \cite{Sorkin:2007qh,spacetimeatoms,dowker_2020}.\footnote{The passage of time in Newtonian mechanics is tied to change. Time in GR is bound up in the precedence order.  If we look forward to a successful theory of quantum gravity in which the passage of time is a birth process, then the history of the concept of time in physics might be represented by a trio of aphorisms: Time in the Past was Change; Time in the Present is Order; Time in the Future will be Process. However, considering Th. Stcherbatsky's account of early Buddhist thought, ``This constitutes the second characteristic feature of early Buddhism: no Matter, no Substance, only separate elements, momentary flashes of efficient energy without any substance in them, perpetual becoming, a flow of existential moments'' (page 4-5 of \cite{Stcherbatsky:1932}) and the views of the Vibhajavadin school (for example page 43 of \cite{Stcherbatsky:1922}), maybe this trio should better be ordered in time as Process$\prec$Change$\prec$Order$\prec$Process.}  The article will not rely on the details of these specific models and we take from the models only the concept of a partially ordered birth process. The dynamical hypothesis is:\\
\noindent  In the yet-to-be-discovered physical dynamics of causal set theory,  events such as $E_{fire}$ and the NCC-events of the next section \ref{sec:consciousness}  occur in a partially ordered birth process. \\

\noindent In the  birth process there are three distinct concepts pertaining to events: \\
(a) an event-as-such $E$, (b) the process of an event $E$ occurring and (c) an event $E$ that has occurred. I expand on  this three-way distinction using the example of the explosion event $E_{fire}$ in CST discussed in the previous subsection 
\ref{sec:cstevents}. \\

\noindent (a) An event-as-such is what I have been referring to hitherto simply as an event, for example $E_{fire}$. David Reid has suggested the term ``occurable'' as a synonym for the term event or event-as-such that captures the sense that I want to emphasise.  
The event $E_{fire}$ may or may not occur, the firecracker may or may not explode. \\ 

\noindent (b) The process of $E_{fire}$ occurring, or the \textit{occurrence} of $E_{fire}$ for short, is the  process of the birth of the $10^{144}$ or so atomic-events that compose $E_{fire}$, in their  order of precedence. Referring to figure \ref{fig:mink}, recall that the precedence order of the example atomic-events shown is  $o\prec r$, $r\prec s$, $o\prec s$, $p\prec q$, $p\prec s$. The physical order of the births of the atomic-events of $E_{fire}$ including $o,p,q,r$ and $s$ is a partial order. \\

\noindent (c) An event that has occurred is an event all of whose atomic-sub-events have been born. To distinguish (a) and (c) I will use the term ``occurred-event'' to refer to an event that has occurred. The same notation $E$ is used both for the event-as-such and for the occurred-event but in the latter case I will refer to it as ``the occurred-event $E$''. 

The possibly unfamiliar context of causal sets may make the above discussion of and distinction between the three concepts (a), (b) and (c) seem more obscure than it is. The very same three concepts exist for any stochastic process. Consider for example the random walk introduced in section \ref{sec:events} and the dynamic physical theory A. $E(5;13)$ is an event-as-such, an occurable that may or may not happen.  As already stated, in the dynamic process the occurrence of $E(5;13)$ is the stepping of the walker to the integer $5$ (from $6$ or $4$) at stage $13$ of the process. The event $E(5;13)$ and its occurrence are different concepts. And then, once stage $13$ in the process has been reached, if the walker stepped onto position $5$ at stage $13$, then $E(5;13)$ is an occurred-event (c). 

\subsection{No objective external picture of the birth process}\label{sec:noEOP}

Let us try to conceive of the event $E_{fire}$ occurring in the dynamic process.  For $E_{fire}$ to occur each of its atoms must be born, including $s$ for example, referring to figure \ref{fig:mink}.  $s$ is preceded by both $p$ and $r$ and so  $p$ and $r$ must both be born before $s$ can be born in the process. There is no order in which $p$ and $r$ are born: $p$ is not born before $r$ and $r$ is not born before $p$. The only way to conceive of the births of $r$ and $p$ dynamically occurring seems to be to conceive of them being born simultaneously and that would be an unphysical thought. 
Moreover, even if one were to toy with thinking of the births of  $r$ and $p$ as simultaneous, 
 one would run into a downright contradiction for the same argument can be made for 
$p$ and $o$. But if the births of $p$ and $o$  are also simultaneous then the births of $r$ and $o$ are simultaneous which contradicts $o\prec r$.  There is no external objective picture  of the process happening. 
Neither is there an external objective picture  of the world-as-it-is-momentarily during the  process. For, if there were, then the sequence of transitions from each external objective picture  of the world-as-it-is-momentarily to a subsequent external objective picture  of the world-as-it-is-momentarily would be an external objective picture of the process. 

A consequence is that the distinction between (a) events-as-such and (c) occurred events is not an external objective distinction: there is no external objective picture of which events-as-such have become occurred-events in the process, as the process is happening.

\subsection{A partly subjective external picture of the process}\label{sec:ESP}

We, as scientists using the theory, can have an external  picture of the process happening by using labels but that external picture is necessarily partly subjective, partly unphysical. 
  
  The $n \approx 10^{144}$ atomic-events in $E_{fire}$ can be labelled by the  numbers, $\{0, 1, 2, \dots
 10^{144}\}$ in such a way that the labels respect the atomic-events' precedence order: if the atom labelled $n$ precedes the atom labelled $m$ then $n<m$. 
 Such a labelling is called a natural labelling. There are many natural labellings of the same partial order. Given a natural labelling, one can 
 conceive of the process dynamically as the \textit{sequential} process of the birth of the  atoms, one at a time, in the total order of the labelling: $0$ is born and then $1$ and then $2$ etc.  The label-induced picture of the process is a sequence of stages labelled by $n$: at stage $n$ the atom labelled by $n$ is born.
The label-induced  picture of the process is partly subjective because 
the label-order exists in the minds of the scientists using the theory but the label-order of two
atomic-events that have no precedence order is not physical.  The picture is also partly objective because the label-order of 
two atomic-events that do have a  precedence order is physical. A choice of natural labelling is a choice of {gauge}. 

We can also use a natural labelling in a cosmological context 
to give a partly subjective picture of the whole growing universe during the process. Say the universe being built up 
in the process is an event with a causet substrate whose continuum approximation is part of Minkowski spacetime  after  a ``Beginning'' at coordinate $x^0 = - 10^{16} s$: a Big Bang without the Bang. 
The atomic-events can be given a natural labelling by the natural numbers $\{0, 1, 2, \dots\}$. Suppose a natural  labelling is chosen. For each $n$, consider the finite event $E_n$ whose substrate is the subcauset labelled by the numbers $\{0,1,2,\dots n\}$. $E_n$ is composed of $n+1$ atomic-events. Then the growing world during the label-induced sequential process is the nested sequence of events $\{E_n\}$. A different natural labelling of the same physical process will give a different partly subjective picture of the growing world. 

In the external, partly subjective picture of the process provided by a particular natural labelling, the distinction between (a) events-as-such and (c) occurred-events can be made at each stage of the process, as the process is happening. At stage $n+1$ in the label-induced process,  every subevent of $E_n$ is an occurred-event.  
 
Scientists need to use such a partly subjective picture because we need an external picture of the process that we can think about, work with and communicate to one another. Indeed, the Classical Sequential Growth models from which the concept of the partially ordered birth process is taken are defined using natural labellings \cite{Rideout:1999ub}.  
  
\subsection{Comparison with the random walk} 
It may be helpful at this point to recall the random walk discussed in section \ref{sec:events} and to compare the birth process in CST and the dynamic process of the random walker's stepping. 
Section \ref{sec:events} ended with the Note that the dynamic process of the random walk admits an external view of the process: for example a movie of the walker stepping left or right in a sequence of steps. The point is that in the random walk the ``atomic-events'' are the single steps and they are physically, objectively linearly ordered so the order of the sequence of the steps in the movie---and/or in the minds of physicists as they think about and work with the process---is  physical and objective.  A movie of a random walker is an external objective view of the dynamic process as it happens.  In the birth process in CST however, as we have seen, the atomic-events are physically, objectively partially ordered and the physical partial order is what precludes  an external objective view of the dynamic birth process, even one which is itself dynamic like a movie.  

\subsection{An objective internal view}\label{sec:OIV}

An objective external view of the birth process does not exist. But, according to the Proposal, an objective view is possible if the view is \textit{internal}.   One can view the process objectively if one views it from the \textit{inside}, if one is part of the physics. 
Furthermore, from the inside the distinction between (a) events-as-such and (c) occurred-events can be made objectively. \\

\noindent Live experience, to which we now turn,  is an objective internal view of the birth process.

\section{The Hard Problem of consciousness in causal set theory} \label{sec:consciousness} 

Let us put the tool of the partially ordered birth process in CST to work on the Hard Problem.
I assume there is an NCC theory based on certain events, NCC-events.  The Proposal  does the same job whatever the physical correlates of conscious experiences in the NCC theory are and is neutral on whether they  are confined to the brain, or involve other parts of the body, or are events in a computer, or whatever.  But for the Proposal  to be useful, it is necessary that the problem of identifying NCC-events and constructing an NCC theory---the problem that, I believe, Anil Seth calls the Real Problem of Consciousness \cite{seth2021being}---can be solved. Let $E_{exp}$, then,  be an NCC-event that is the correlate, in the NCC theory, of a particular conscious experience of a subject $A$.

The hypotheses of subsections \ref{sec:cstevents} and \ref{sec:dynevents} applied specifically to $E_{exp}$ are that the NCC-event $E_{exp}$ in CST has a causet substrate that recovers a portion of Minkowski spacetime and that $E_{exp}$ occurs in CST as the process of the partially ordered birth of the atomic-events that compose $E_{exp}$.  The Proposal  stated in the introduction section \ref{sec:intro} can then be restated in a slightly more concrete way:

\noindent The birth of the atomic-events that compose $E_{exp}$ in CST correlates with $A$ \textit{having} that conscious experience live, in real time.\\

\noindent A corollary of the Proposal  is that the occurred-event $E_{exp}$ correlates with $A$ having had that experience. 

\subsection{Qualities of conscious experience} \label{sec:qualities} 

If we accept that the perception of time passing is a certain quality of conscious experience and not any particular content of conscious experience, then confidence in the Proposal  can be built by intuitively connecting properties of the partially ordered birth process to qualities of conscious experience. 
 
 \begin{itemize}

\item [(1)] \textit{In the birth process each atomic-event that is born, is born once. The birth of an atomic-event is momentary. Immediately an atomic-event is born, it becomes part of the past.}

\noindent Conscious experience has been described as momentary, fleeting and of the now. \\

\item [(2)] \textit{The birth process is unceasing. }

\noindent Conscious experience of time passing has been described as inexorable. \\

\item[(3)]  \textit{There is no external objective picture of the birth process. The birth process can only be viewed from inside the world, as it happens.}

\noindent Conscious experience has been described as live. \\

\item[(4)] \textit{The birth process can only be viewed objectively from inside the world. The entity having a live experience in the world cannot copy and communicate that live experience, cannot share it with another entity since to do so would be to create a picture of the process that is external and objective.}

\noindent Conscious experience has been described as internal and private.\\

\item[(5)] \textit{The birth process is objective.}

\noindent Conscious experience has been described as indubitable. \\

\item[(6)] \textit{The process of the birth of the atomic-events that compose $E_{exp}$ correlates with the live experience.}

\noindent Conscious experience has been described as immediate (un-mediated).

\end{itemize}

\subsection{In the light of the Proposal  }\label{sec:inlightof} 

\subsubsection{Consciousness in the Block}

If the birth process were to run to its infinite completion, the result would be one single vast inextendible occurred-event, an occurred-Block Universe composed of all the atomic-events that were born in the process, and in which every subevent is an occurred-event.  The universe would have run its course. Nothing occurs any more and so there would be no live experiences in such a world {because} nothing occurs. There is consciousness in this occurred-Block Universe only in the sense that there \textit{were} entities that \textit{had} conscious experiences. The human beings, say,  with their whole histories, including histories of their brains, neurons and all, incorporated in such an occurred-Block Universe are not zombies as usually conceived but they do not have live experiences because they do not have experiences \textit{any more}. They had experiences but those experiences are over.
In the light of the Proposal, therefore, my answer to the 
question, ``What would it look like if it looked as if our world were a Block Universe?"  is, ``It wouldn't look like anything because it would all be over."\footnote{I thank Henrique Gomes for asking this question after my talk at the ``Time in Cosmology'' meeting at Perimeter Institute, Waterloo ON, Canada, June 2016.}

\subsubsection{What breathes fire into the theory}

The NCC theory is a theory based on concepts of NCC-events-as-such. 
Live experience is the occurrence of NCC-events such as $E_{exp}$ and the occurrence of an NCC-event is the partially ordered birth of its atomic-events. What is missing from the NCC theory, what breathes fire into it, is the birth process.

\subsubsection[The knowledge argument]{The knowledge argument {\normalfont{\cite{sep-qualia-knowledge}}} }
 
Anyone who knows the NCC theory knows the full physical account of the event-as-such $E_{exp}$ that is ``$A$ had the conscious experience of seeing a red ball''. But one cannot 
 know the physical correlate of $A$ having the experience, live and in real time, because 
the correlate of \textit{that} is the partially ordered birth process.  
There is no objective, external picture of the birth process, as its true dynamic self, that can be communicated, shared or known. Even relaxing Einstein's stipulation that theoretical concepts be ``laid down in books'' to include theoretical concepts that can be captured in movies, such as the dynamic random walk process, does not avail us here, because of the partially ordered nature of the birth process.  \\
\noindent  Only the subject $A$ can view the occurrence of event $E_{exp}$, experience it, from within the process, live and as it happens.

\subsubsection{How consciousness interacts with spacetime, particles and fields}

Conscious experience is the physical process of the birth of the atomic-events that comprise NCC-events. The objective, ubiquitous birth process is the creation of physical stuff,  physical \textit{4-dimensional spacetime stuff},  particle trajectories and spacetime field configurations together with their spacetime substrate. It makes as much sense to ask how consciousness interacts with physical stuff as to ask how the birth of a baby interacts with a baby. Without the process there is no physical stuff and the physical stuff is what the process creates. 

\subsubsection[Monism]{Monism{\normalfont{ \cite{sep-neutral-monism}}}} 
 
The framing of the partially ordered birth process in this article denies monism by including \textit{both} the atomic-events \textit{and} the process of the occurrence of atomic-events as physical elements of CST. An event and the process of the occurrence of an event are completely different types of physical thing: the birth of a baby is not a baby, the birth of a spacetime atom is not a spacetime atom, the occurrence of an event is not an event. This is the position taken in \cite{dowker_2020} and it is inherent in the very terms  becoming  and birth. The concepts of becoming and birth presuppose that there is something coming into being, something being born.  
 
Another possibility, however, is that the physics consists only of the process. Could the physics be just, ``atomic-events occur ceaselessly  in a partial order of precedence,'' and nothing more (one would have to drop the birth metaphor and the concept of becoming)?  If this position could be developed within CST it would be the sort of monistic physics that Galen Strawson is gesturing towards when he writes ``All things [...] are made of the same single kind of fundamental stuff:
material or physical stuff. We can put it in more dynamic terms, because stuff is best
thought of as process, process-stuff: everything in the universe, conscious or not, is
wholly a matter of physical process, physical goings-on.'' \cite{Strawson2021-STROYM}  

\subsubsection[Panpsychism]{Panpsychism{\normalfont{ \cite{sep-panpsychism}}}} 

The  Proposal  is sympathetic to panpsychism  only to the extent that the birth process is universal to the whole physical world. The question, ``Which entities have conscious experience?'' 
is the question, ``Which events are NCC-events?'' and defines the quest for an NCC theory. 
For example, a supernova is an event and is composed of atomic-events. The process of the partially ordered birth of the atomic-events composing the supernova-event  is the occurrence of the supernova-event. 
The question ``Was the supernova that occurred conscious?'' is a question that the NCC theory should answer.
   
\subsubsection{Fundamentalism about consciousness}

The partially ordered birth process is fundamental and not emergent in causal set theory. ``Time'' in the sense of Becoming is fundamental and not emergent in causal set theory. The birth process cannot be recovered from anything more basic. The NCC theory on the other hand will be constructed using emergent concepts in biology, neuroscience and cognitive science such as neurons, information processing and superfast model fitting.

\subsection{Cortes, Smolin and Verde} \label{sec:sv}
  In this subsection I compare and contrast the Proposal  to the proposal of 
Cortes, Smolin and Verde (CSV) in  \cite{pittphilsci19530}.
CSV suggest that the Hard Problem of consciousness, the problem of the physicality of the passage of time and the problem of quantum gravity have a common solution. This is very much in the 
same spirit as the Proposal, as is the concept they refer to as  \textit{resolving time} (pages 4-5 of \cite{pittphilsci19530}): \\
\noindent ``In  (Smolin \& Verde, 2021) we define an active conception of time as
the process that resolves indefinite circumstances to definite. We call the version of time
we champion \textit{resolving time}. It is what Bergson called \textit{creative time} (Bergson,??) [sic]. The 
world is literally remade over and over again, event by event, by the work of active
time.''

To the extent that CSV's concept of resolving time is a creative {process} it has sympathy with the 
partially ordered birth process of CST. And, to the extent that the process of resolving time is what CSV call Mode II  physics then their proposal, ``Mode II is where we will find the physical correlates of consciousness'' also has 
sympathy with the Proposal. 

There are differences between CSV's framework and the physics of partially ordered birth in CST, illustrated in this passage   \cite{pittphilsci19530}:\\
\noindent ``An event is a process, each of which does its job of resolution,
then vanishes as it creates a few of the next generation. Then, its work being done, it is
gone. A time created world is not a four dimensional manifold, it is a continually
recreated, roughly three dimensional network of processes.''\\
In CST an event is a distinct concept from the process, so CSV use the terms differently in order to say ``an event is a process''.
The biggest conceptual difference between CSV and the CST birth process, however,  seems to me to be  the concept of the ``time created world'', ``a continually recreated, roughly three dimensional network of processes''. A 3D network and a sequence of different 3D networks seem to me to be external, objective pictures.  In the CST birth process, there is no physical concept that is three dimensional. 
 
The phrase ``network of processes'' indicates that CSV's time created world is discrete and this is another point of sympathy with the Proposal. Is discreteness really necessary to the Proposal, though? Here I just mention Section 8 of \cite{Sorkin:2007qh}  which is a commentary on the question of discreteness vs continuity in the context of becoming, and also  Dieks \cite{Dieks} who comes close to proposing a partially ordered becoming process for continuum spacetime, but messes with the proposal by claiming ``There is no need to augment the block universe in any way'',  which claim I deny.  

\section{Discussion}\label{sec:discussion}

The Proposal  will be judged on whether it rings true but more importantly on its usefulness in the production of new scientific knowledge. In the study of consciousness, the Proposal  may indicate or illuminate aspects of our conscious experience that  have hitherto gone unnoticed and unreported.  Conscious experiences may be re-interpreted in the light of the Proposal. The Proposal  may therefore help to improve the theoretical concepts in NCC theories and help to improve the reliability of the self-reporting of conscious experiences that is used as data in the development and testing of NCC theories. 

Turning things around, if we judge that the concept of the partially ordered occurrence of atomic-events in CST is useful for understanding our experience, then we will want that concept to be part of the further development of CST as a theory of quantum gravity. In order to apply the birth process in CST to the Hard Problem in section \ref{sec:consciousness} I  hypothesized that an NCC-event $E_{exp}$ in CST is composed of atomic-events that are born in their partial order of precedence. This hypothesis begs many questions, especially questions about quantum theory. How does CST recover the Standard Model of particle physics in Minkowski spacetime, a quantum theory of matter that is well-established and that, most likely, recovers the physics, chemistry and biology used in working out the NCC theory? What about the quantum nature and dynamics of the causet substrate itself? How will that be described in CST? Answering these and other questions is work in progress in CST and we may choose to let the Proposal  guide us in that work. I give here a very brief sketch of progress in the development of a quantum theory of causal sets.  The framework in which the quantum dynamics of causal sets is being sought is the path integral approach to the foundations of quantum theory.  The two most developed path integral approaches to quantum foundations are the related programmes of  generalised quantum mechanics (GQM) proposed and championed by James B. Hartle \cite{Hartle:1992as, Hartle:1998yg, Hartle:2006nx} and quantum measure theory  (QMT) proposed and championed by Sorkin \cite{Sorkin:1994dt, Sorkin:1995nj, Sorkin:2006wq, Sorkin:2007uc, Sorkin:2010kg}.  In QMT quantum mechanics is explicitly conceived as a species of measure theory,  based on the same concepts of history and event that are basic to GR and to the classical causet growth models. And moreover, in the currently favoured interpretational framework for QMT, every event either occurs or does not occur in the quantum process that generates the physical world, just as in classical measure theory. So far so good. Where QMT diverges from the classical theory is in the pattern of the occurrences of events, their subevents and their combinations, which pattern need not conform to Boolean rules of inference \cite{Sorkin:2007uc, SorkinTetralemma, Incomputable}. The world that is being built up by a quantum process, then, may not be conceivable as an event with a single causet as its substrate but may be something like a causet with ``classically contradictory'' properties. It is possible that in the yet-to-be discovered quantum rules of inference an event occurs but some of its component atomic-events do not occur, or vice versa \cite{Sorkin:2007uc}. I for one 
look forward to wrestling with the implications of such ``anhomomorphic logic'' for our understanding of quantum gravity and of consciousness.\footnote{Sorkin asks what experiences led Buddhist logicians, who did not have access to the technology that enabled the development of quantum mechanics, to the anhomomorphic logic of the Tetralemma \cite{SorkinTetralemma}} 

Returning to the present and to Einstein's epistemology,  the partially ordered birth process in CST blurs the boundary between the world of ideas and the totality of sense-experiences.  The birth process in CST as its live, dynamic self cannot be situated fully in the world of objective, communicable physical concepts {because it is partially ordered}.
 Though the process is a concept in the theory, it can only be apprehended in its dynamic aspect in the manifold of sense-experiences. It can only be viewed fully objectively from within. It has to be lived, it has to be experienced to be apprehended. One cannot know, from the outside, \textit{what it is like}.

\section{Acknowledgments}

I thank Emma Albertini, Jeremy Butterfield, Arad Nasiri, Huw Price, David Reid, Lee Smolin, Rafael Sorkin, Clelia Verde and Yasaman Yazdi for helpful input and discussions. I thank Domenico Giulini for improving the English translation of the passage from Einstein's Autobiographical Notes. Thanks are also due to the Quantum Gravity group at Perimeter Institute, the participants of the online Virtual Causet Seminar and the participants of the WE-Heraeus-Seminar ``Time and Clocks''  for stimulating discussions.

This work was supported in part by STFC grant ST/T000791/1. It was also supported in part by Perimeter Institute for Theoretical Physics. Research at Perimeter Institute is supported by the Government of Canada through Industry Canada and by the Province of Ontario through the Ministry of Economic Development
and Innovation.

\bibliography{../Bibliography/refs}
\bibliographystyle{../Bibliography/jheppub}

\end{document}